\documentclass[conference]{IEEEtran}

\usepackage{standalone}
\usepackage{tikz}
\usetikzlibrary{arrows.meta, shapes}
\providecommand{\keywords}[1]{\textbf{\textit{Keywords---}} #1}

\usepackage{todonotes}
\usepackage{verbatim}

\newcommand{\ie}{\textit{i}.\textit{e}.\ }
\newcommand{\eg}{\textit{e}.\textit{g}.\ }

\usepackage{graphicx}

\usepackage{cite}

\usepackage{url}

\usepackage{balance}

\usepackage{csquotes}

\usepackage{color}

\begin{document}

\title{Lightweight synchronization algorithm with self-calibration for Industrial LoRa Sensor Networks}


\author{\IEEEauthorblockN{Luca Tessaro\IEEEauthorrefmark{1},
Cristiano Raffaldi\IEEEauthorrefmark{2},
Maurizio Rossi\IEEEauthorrefmark{1} and 
Davide Brunelli\IEEEauthorrefmark{1}}

\IEEEauthorblockA{\IEEEauthorrefmark{1}Department of Industrial Engineering, 
University of Trento,
Trento, Italy\\ Email: luca.tessaro-1@alumni.unitn.it, \{davide.brunelli, maurizio.rossi\}@unitn.it}
\IEEEauthorblockA{\IEEEauthorrefmark{2}Adige S.P.A., BLM Group, Levico Terme, Italy\\
Email: rd.team@adige.it}}


\maketitle

\begin{abstract}
Wireless sensor and actuator networks are gaining momentum in the era of Industrial Internet of Things IIoT. 
The usage of the close-loop data from sensors in the manufacturing chain is extending the common monitoring scenario of the Wireless Sensors Networks WSN where data were just logged. 
In this paper we present an accurate timing synchronization for TDMA implemented on the state of art IoT radio, such 
as LoRa, that is a good solution in industrial environments for its high robustness. 
Experimental results show how it is possible to modulate the drift correction and keep the synchronization error within the requirements. 
\end{abstract}


\IEEEpeerreviewmaketitle

\section{Introduction}
\label{sec:introduction}

Synchronization in Wireless Sensor Networks (WSN) can be considered a critical and fundamental requirement to properly correlate the timestamped data acquired by each sensing device with a common reference clock.
Multiple synchronization protocols have been developed over the years to optimize the network for different requirements. 
Among all, Time Division Multiple Access (TDMA), as a channel access method, requires the synchronization as a necessary condition to avoid overlapping messages and a decreasing in network performance.

In this work a lightweight synchronization protocol with an online clock calibration method is presented.
The protocol design is part of a larger industrial project that aims to build a star network to acquire information about some slowly varying parameters around and inside a laser cutting machine.
The machine is specialized in tube cutting and can reach over 20m in length---along the tube axis---which influences the technology choice for a single-hop topology.
Furthermore, the nodes must be battery powered and a low power technology is required.
Given the requirements, the LoRa radio technology from Semtech has been chosen and tested in the environment 
to optimize the radio configuration.
LoRa is a spread spectrum modulation technique 
designed for low power and long range applications (up to 15km in line of sight), but 
its robustness to noise, regardless of the very low throughput (sufficient for the task), makes it suitable for the specific application.

Many different synchronization algorithms are available in literature. 
The proposed one is inspired by the Continuous Clock Synchronization (CCS) algorithm~\cite{CCS} and the modified Dynamic CCS (DCCS)~\cite{DCCS}.
The resulting maximum synchronization error is of few milliseconds due to hardware limitations.
Broadcast synchronization messages are used but, differently from the Reference Broadcast Synchronization (RBS) protocol, here no message exchange between end nodes takes place to reduce the overall number of packets.
Multiple solutions were proposed to correct the unavoidable clock drift in real systems.
Clock skew is mostly due to temperature changes and ageing but other factors influence it with different magnitude (\eg hardware used, circuit design, environmental factors, etc.).
Several studies~\cite{tempcalib1,tempcalib2,tempcalib25,tempcalib3} focus on a temperature based calibration which does not solve all the clock skew problems.
The proposed method uses hardware functionalities of available clock to compensate skew without relying on any sensor.
After the design of the protocol, performance is assessed with some nodes built upon the ST-NUCLEO64 development board with the STM32L073 microcontroller and the Semtech SX1276MB1MAS LoRa radio.

The proposed solution can improve the reliability not only in the specific case, but also in several scenarios where distributed monitoring of a number of sensors spread in a critical environment should be realized, e.g. from the monitoring inside datacenters~\cite{serverFarms},  to solutions that can leverage energy harvesting and sustainability of the nodes in a network\cite{comparison}. 

The paper is organized as follows: Section~\ref{sec:methods} describes the algorithm and motivations of the proposed work, while in Section~\ref{sec:results} experimental results of a real hardware implementation are discussed.
Lastly, conclusions are reported in Section~\ref{sec:conclusion}.

\section{The algorithm design}
\label{sec:methods}

\begin{figure}[!t]
\centering
\includegraphics[width= \textwidth/2]{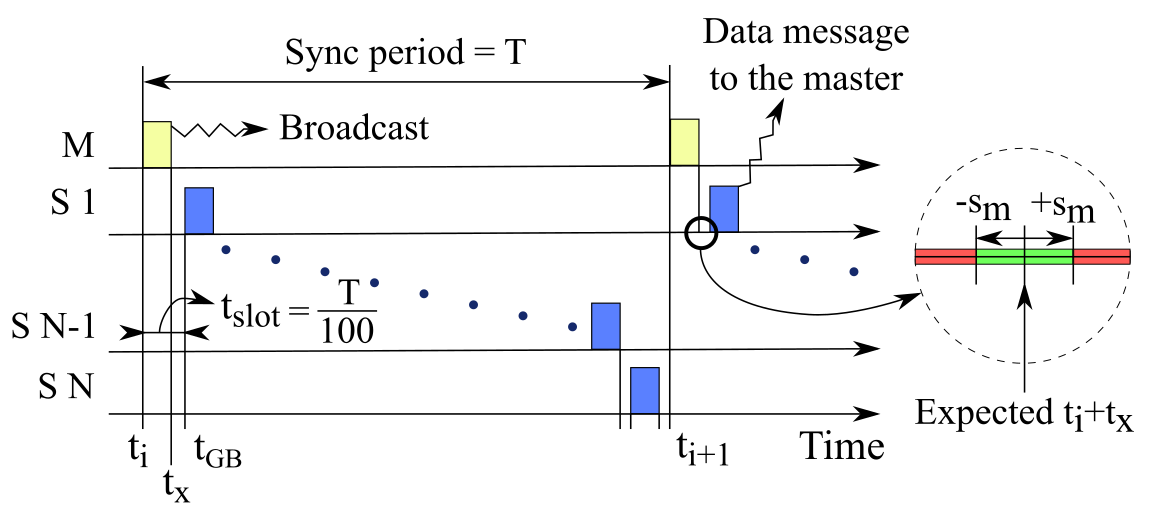}
\caption{Schematic representation of the TDMA protocol}
\label{fig:tdma_schematics}
\end{figure}

The LoRa technology offers a very compelling mix of long range, low power consumption and secure data transmission thus,
as first step of any design, some 
specific configuration tailored to the target application must be performed.
The 
most relevant configuration parameters are:
\begin{itemize}
\item the Spreading Factor (SF) which controls the dilatation in time of the message;
\item the bandwidth (BW) is the width of the channel used (which influences also the time needed to transmit a packet);
\item the Cyclic Redundancy Code (CRC) which helps the receiver to understand if the message is correct or has been corrupted (stronger CRC increases the packet length and consequently the transmission time).
\end{itemize}
Other constant parameters are dictated by the technology itself:
\begin{itemize}
\item LoRa works in the sub-GHz ISM band thus a region dependent frequency is used: different disturbances may be present in different locations or there may be region dependent regulations (\eg this work has been done in EU where there is 1\% transmission duty cycle limitation over 1h for each node).
\item the maximum packet length is 255~Bytes.
\item in addition to the payload, each packet comprehends a preamble and an header which must be considered as overhead.
\end{itemize}
These parameters influence the network design under multiple points of view: transmission time limited by the duty cycle, maximum throughput achievable, maximum packets frequency, etc.
By means of an extensive set of experiments (not reported for the sake of summary and out of the scope of the present work) the optimal LoRa configuration chosen is SF~7, BW~500kHz and CRC~$4/5$.
The self-synchronizing protocol (summarized in Figure~\ref{fig:tdma_schematics}) was designed on top of this radio configuration and a TDMA approach, to optimize the available duty cycle and consequently maximize the throughput.

The lightweight synchronization protocol exploits a periodic broadcast message sent by the master node (M in Figure~\ref{fig:tdma_schematics}) in every period T, which contains the absolute reference timestamp.
Any end device $(S_1, \ldots, S_{N-1}, S_N)$ 
is in charge of its own synchronization, allowing the protocol to scale-up to an unlimited number of end devices.
In the specific case of LoRa technology, each node can use 1\% maximum of the period to maximize the throughput, which means maximum 100 nodes can join the network. 
To increase the number of allowed nodes it is necessary to reduce the single node duty cycle (consequently the data payload).
Notice that the period T is not influencing the other tasks that can have their own frequency (\eg the acquisition is performed at 1Hz).

Each node updates the local timestamp according to the received one, 
adjusted considering some constants: (i) the preparation time of the message at the master side, (ii) the transmission time of the packet (the time occurred between transmission start and transmission end) and (iii) the elaboration time of the receiver;
summarized in constant $t_x$.
Notice that the 
time-of-flight is considered negligible. 
With the formula provided by Semtech~\cite{sx1276datasheet}, 
verifyed with experiments, the transmission time 
for a 255B payload is of about 100ms which needs to coincide with $t_{slot}$ to maximize throughput (thus $T = 100\cdot t_{slot} = 10 s$).
As a requirement, the protocol has a 10B header and needs to fit data from nodes equipped with 11 sensors (16bit resolution).
The total payload length of 230B is shorter than the maximum, allowing to introduce a 9~ms guard band (GB) to takle unavoidable synchronization errors in the order of $\pm$4.5~ms ($t_{GB}$).
As a result, the protocol keeps the LoRa network synchronized with a maximum clock drift of 4.5~ms every 10~s.

In a real scenario, with low cost hardware and in harsh environments the clock drift can be larger than the guard band.
In this specific case, the on-board RTC oscillator 
 responsible for the drift is used to achieve self-calibration.
Multiple different strategies can be implemented to compensate this error.
A first solution to the problem is to create a virtual software clock that elaborates the hardware clock information to compensate for the skew and the offset. 
This method needs an estimation of the error and clock skew to work properly and may require too many computational resources.
A simpler solution is to use the hardware calibration functionalities to manipulate the clock skew \cite{WuSPmagazine}.
In the case under study, the clock skew is controlled through the use of some calibration values associated to dedicated registers included in the hardware. 
This allows to increase or reduce the clock frequency \ie the drift can be controlled.
Usually, this calibration value is computed once in static tests by comparing the oscillator signal to a more precise frequency source.
This solution permits an average drift correction but the variable effects such as temperature changes, power supply variations or ageing are still present.
To compensate for temperature variations a lookup table or a mathematical relation can be used to relate changes to clock calibration correction \cite{tempcalib2}.
This method alone does not solve the problems due to ageing or power supply changes. 
Furthermore, each device has slightly different hardware and may require a different calibration table.
In this work, the reported factors are compensated even if the algorithm does not need any information about the device status. 
Indeed, only the information about previous synchronization messages is needed.
The resulting runtime calibration corrects also for unknown sources of drift.

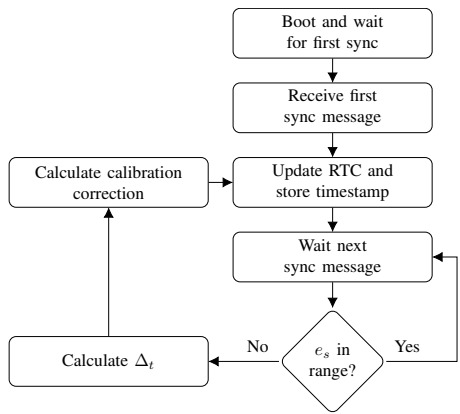
\begin{figure}[!t]
\centering
    
\tikzset{%
  >={Latex[width=2mm,length=2mm]},
            base/.style = {rectangle, rounded corners, draw=black,
                           minimum width=4cm, minimum height=1cm,
                           text centered},
        decision/.style = {base, diamond, minimum width=1cm, minimum height=0.5cm, node distance=2.1cm},
       startstop/.style = {base, fill=red!30},
    activityRuns/.style = {base, fill=green!30},
         process/.style = {base, minimum width=2.5cm, fill=orange!15},
}
\begin{tikzpicture}[node distance=1.5cm,
    every node/.style={fill=white}, align=center]
  \node (boot)            [base]                  {Boot and wait\\ for first sync};
  \node (setRTC)          [base, below of=boot]   {Receive first\\ sync message};
  \node (updateRTC)       [base, below of=setRTC] {Update RTC and\\ store timestamp};
  \node (wait)            [base, below of=updateRTC] {Wait next\\ sync message};
  \node (checkError)      [decision, below of=wait] {$e_s$ in\\ range?};
  
  \node (yes)             [right of=checkError, yshift=0.3cm]  {Yes};
  \node (no)              [left of=checkError, yshift=0.3cm]     {No};
  \node (calculateDelta)  [base, left of=checkError, node distance=4.5cm ]      {Calculate $\Delta_t$};
  \node (calculateCalib) [base, above of=calculateDelta, yshift=2.1cm] {Calculate calibration\\ correction};
  \draw[->]             (boot) -- (setRTC);
  \draw[->]           (setRTC) -- (updateRTC);
  \draw[->]        (updateRTC) -- (wait);
  \draw[->]             (wait) -- (checkError);
  \draw[->]       (checkError) -- ++(2.5,0) -- ++(0,2.1) -- (wait);
  \draw[->]       (checkError) -- (calculateDelta);
  \draw[->]   (calculateDelta) -- (calculateCalib);
  \draw[->]   (calculateCalib) -- (updateRTC);

  \end{tikzpicture}
\caption{Online calibration algorithm work-flow.}
\label{fig:workflow}
\end{figure}

The algorithm steps reported in Figure \ref{fig:workflow} can be synthesized as follows, where the i index represents the i-th synchronization period that starts at $t_i$ (Figure \ref{fig:tdma_schematics}):
\begin{itemize}
\item a constant error margin is chosen. This margin indicates the maximum acceptable synchronization error. In this particular case, with a maximum error of $\pm 4.5$~ms the margin ($s_m$) has been set to 2~ms.
\item when the first synchronization message is received, the RTC calendar registers are written and the timestamp is temporarily stored for future use ($t_{old}$).
\item at the next synchronization message, the received timestamp will be compared with the one available from the RTC registers and the result will be the synchronization error $e_s = t_{sync} - t_{new} = t_i^{master}+t_x - (t_i^{node}+t_x)$, where $t_{new}$ coincides with the time measured by the node at message reception and $t_{sync}$ is the master's timestamp corrected with the transmission time $t_x$.
\item if the error $e_s$ is within the range $[-s_m, +s_m]$, no operation is performed.
\item if the error $e_s$ is outside this range, the error is not acceptable and a correction will be performed. The time since last RTC update is then calculated as $\Delta_t = t_{new} - t_{old} = t_i + t_x - (t_{i-j} + t_x)$, where $j$ is the number of cycles since the last RTC update.
$\Delta_t$ is now used as indication of the magnitude of the oscillator's drift (the smaller the drifting behaviour, the larger $\Delta_t$). $\Delta_t$ and $sign(e_s)$ are then used to calculate the calibration parameter variation to apply.
Once the calibration value is updated, the new timestamp is written into the RTC and $t_{new}$ becomes $t_{old}$.
\end{itemize} 

In the next section some applicative tests of the designed algorithm are reported.

\section{Results}
\label{sec:results}

The hardware used to keep track of the time can reach a maximum time resolution of $1/1024$ of second. 
Consequently, the error 
resolution is also around 1~ms. 
If a smaller error resolution is needed, additional timers with higher time resolution 
must be used. 
Furthermore, we verified that the transmission time calculated using the formula from datasheet \cite{sx1276datasheet} is not reliable for smaller time resolutions and consequently is rounded up to the millisecond.
Also the calibration values used in the RTC registers are discrete and an increment 
corresponds to a clock correction of about 0.955~ppm, as reported in the hardware datasheet (full range 
is from -487.1~ppm to 488.5~ppm). 

Before testing the calibration algorithm, the synchronization error between each message is measured.
The measurement system consists of a pair of nodes (one behaving as master and the other as slave) which are set to pull a digital pin high at $t_i$. 
An oscilloscope 
measures the effective temporal difference of the two rising edges that coincide with the synchronization error.
The first test is performed with a clock that is drifting 
of few ppm, resulting in an error of $0.16 \pm 0.29 ms$. 
Secondly a larger drifting behaviour (430~ppm) is forced and 
results in a larger average synchronization error ($4.54 \pm 1.28 ms$) 
outside the limits of the GB. 
For this reason a calibration is necessary. 

\begin{figure}[!t]
\centering
\includegraphics[width=\textwidth/2, trim=4 0 10 20, clip]{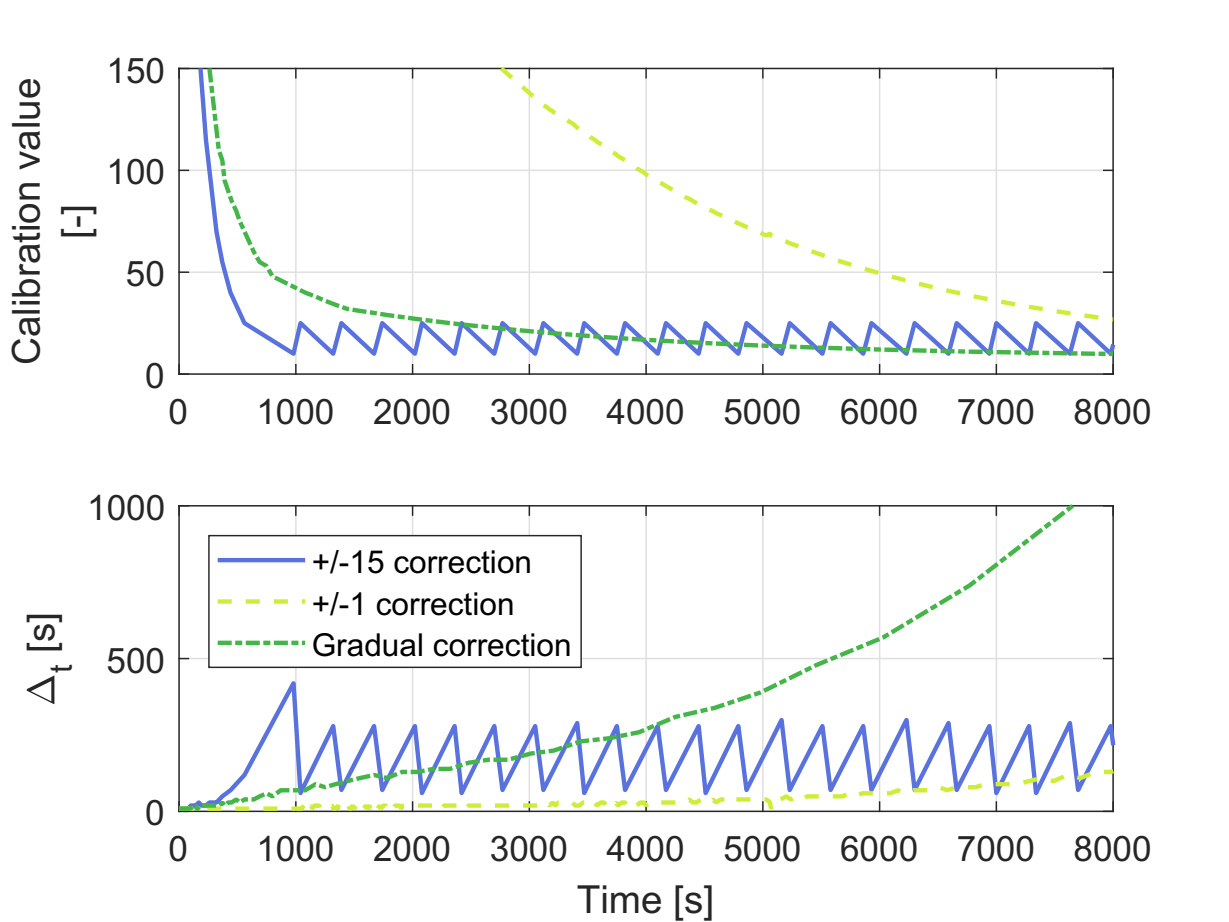}
\caption{Calibration value and $\Delta_t$ for three different correction policies.}
\label{fig:gradualvs10}
\end{figure}

To correct the calibration value over the time, three solutions are tested: the first two apply a constant correction (large with 15~ppm and small with 1~ppm) while the last one uses an adaptive correction, from a lookup table, that depends on 
$\Delta_t$ parameter. 
The 15~ppm solution has a fast convergence rate but in steady state presents some unacceptable oscillations; while the 1~ppm solution shows good performance in steady state but requires longer time (several hours) to converge.
Notice that, even if not shown in the figures, also the 1~ppm solution tends to oscillate at steady state because even with the calibration there is always some drift that, at this point, is below the resolution available to calibrate the clock. 
A simple workaround to avoid the steady state oscillation is to let the calibration value unchanged for $\Delta_t$ greater than a chosen limit value. 
The gradual correction proposed merges the good characteristics of the constant corrections and consists of the lookup table reported in table \ref{tab:lookup}: to different values of $\Delta_t$ correspond different correction values. 
In the specific case, $\Delta_t$ is normalized with the synchronization period to obtain the number of cycles since the last RTC update.
With the non constant correction, the clock converges faster when far from the optimal calibration and keeps the calibration with small changes in steady state (as clearly visible in figure \ref{fig:gradualvs10}).
For every correction, after a first transient behaviour a largely drifting node reaches a steady state in which a reduced synchronization error occurs and thus longer synchronization periods can be used. 
With $\Delta_t$ as only necessary information it is possible to compensate for temperature changes and other effects such as ageing.
Figure \ref{fig:gradualvs10} shows also the increasing trend of $\Delta_t$ which can be seen alternatively as the synchronization period necessary to keep the system in sync.


\begin{figure}[!t]
\centering
\includegraphics[width= \textwidth/2, trim=4 0 10 10, clip]{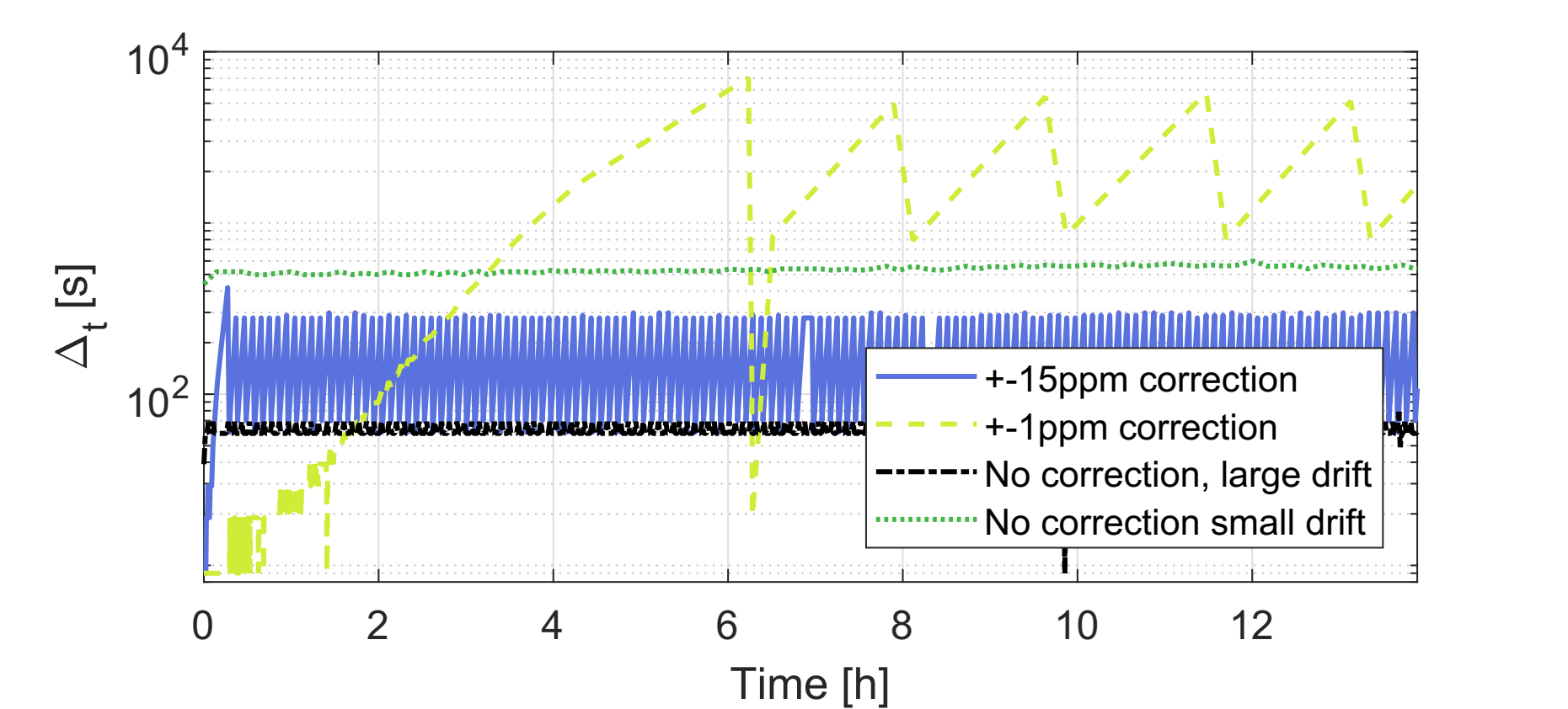}
\caption{The effect of two constant corrections with different magnitude.}
\label{fig:10vs1}
\end{figure}


\begin{table}[t]
    \renewcommand{\arraystretch}{1.3}
    \caption{Gradual correction lookup table}
    \label{tab:lookup}
    \centering
    \begin{tabular}{c||c}
    \hline
    \bfseries Cycles & \bfseries Correction\\
    \hline
    \hline
    $>10$ & 1ppm\\
    \hline
    10, 9, 8, 7, 6 & 2 ppm\\
    \hline
    5, 4, 3 & 5 ppm\\
    \hline
    2, 1 & 10 ppm\\
    \hline
    \end{tabular}
    \end{table}

\begin{figure}[!t]
\centering
\includegraphics[width=\textwidth/2, trim=4 0 10 10, clip]{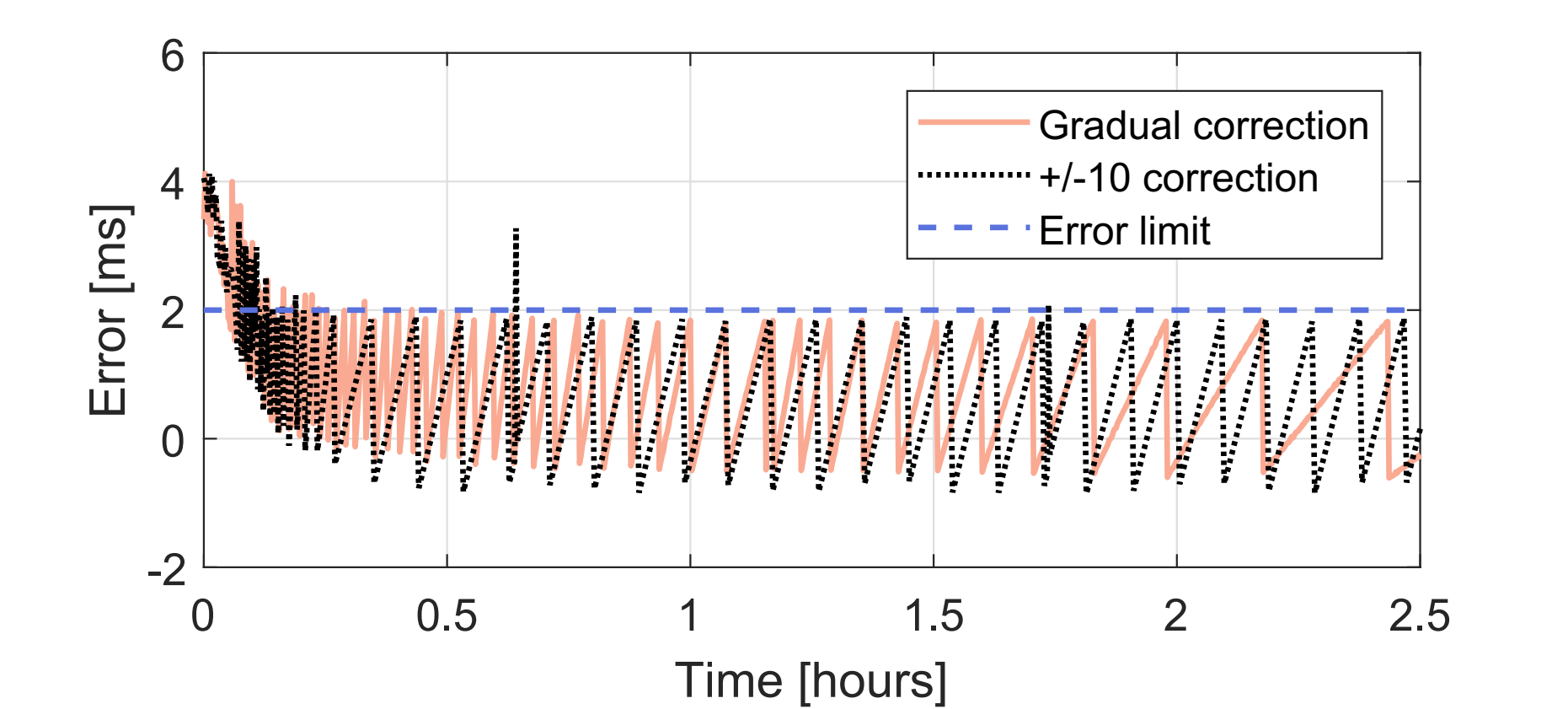}
\caption{Error in time with gradual and fixed $\pm 10$ ppm correction}
\label{fig:errorAndLimit}
\end{figure}
  
Figure \ref{fig:10vs1} reports how the $\Delta_t$ changes in time with a drifting clock. It is clearly visible that a fixed calibration value does not bring any advantage while a correction policy increases $\Delta_t$ until an oscillating behaviours starts as previously explained (here are visible also the oscillations of the 1~ppm correction). The best performance is achieved when the minimum correction of 1~ppm is adopted.
Indeed, for the smallest correction, $\Delta_t$ is in the order of tens of minutes, while for larger corrections it can reach only some minutes.

Figure \ref{fig:errorAndLimit} shows a comparison of the synchronization error between the fixed correction and the gradual correction methods. 
In both cases, at boot the drift in each single period is enough to create a synchronization error greater than the available margin. 
With the calibration, the error decreases until in one single period it doesn't get out of the limits. 
Here the graph starts to resemble a sawtooth because the RTC update brings the error inside the limits and multiple periods are necessary before it exceed this margin again. 
It is also visible that $\Delta_t$ is increasing towards the end of the experiment as the \textit{teeth} become longer and longer. 
Even if at the beginning the behaviour is similar, at steady state the different correction method used causes the system to oscillate with different periods. 
The gradual corrections results into longer periods thus is considered the best solution.

\begin{figure}[!t]
\centering
\includegraphics[width=0.47\textwidth]{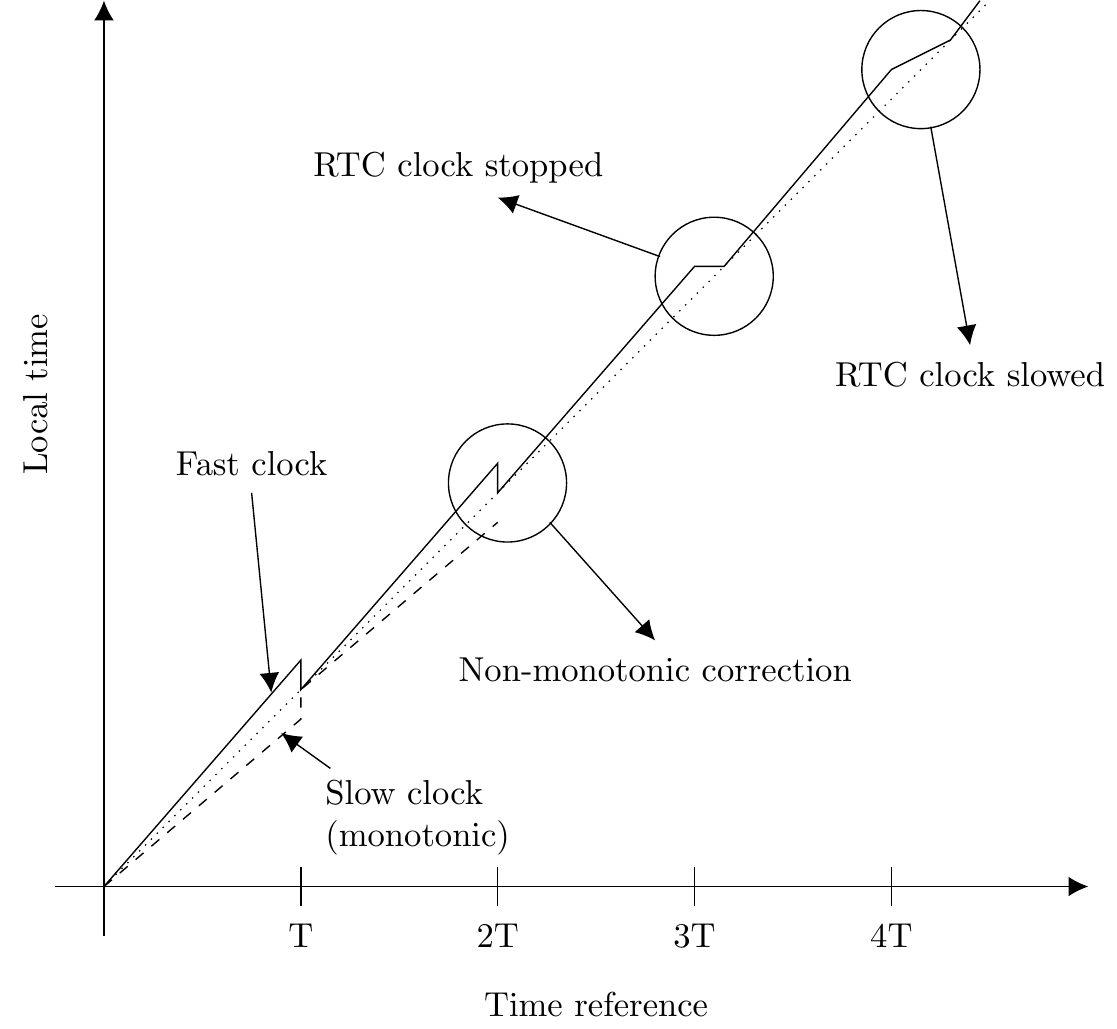}
\caption{The three different time corrections proposed: the issue of the non-monotonicity is solved by stopping or slowing the RTC clock until the counter reaches the correct value.}
\label{fig:monotonic}
\end{figure}

Notice that, with the time register overwrite, the time is not a continuous function and monotonicity is not guaranteed.
Indeed, for a fast clock, during the RTC update it is necessary to insert a time older than the one currently 
present in the counter.
The difference between the two values is $e_s$ and depending on the application it may be or not a problem.
In this specific case, the synchronization protocol is built on top of the TDMA protocol requirements, thus the error is inside the GB and monotonicity is not required.
Also, the application monitors only slow time-varying quantities, thus the 2~ms error is not a problem and the time reference monotonicity at such a small scale is not required.
Nonetheless, to solve this problem, instead of overwriting the counter value it is possible to stop (or slow down) the RTC for the number of pulses necessary to reach the correct time reference.
The two different methods are shown in figure \ref{fig:monotonic}.
The operation needs another timer to correctly measure the time while the RTC behaviour is modified.
This feature is not required for the protocol thus it has not been tested.

\section{Conclusion}
\label{sec:conclusion}

More sophisticated algorithms can be implemented upon this simple implementation. 
It may be interesting to use a small synchronization period at the network startup (when the nodes are not well calibrated) and then after some periods to increase the time between each synchronization to spare duty cycle. With this idea it is possible to increase or decrease the synchronization message frequency dynamically in response to a sudden change in environmental conditions that may cause non constant drift.
As a further step, the algorithm could be tested to see its performance with time resolution of microseconds instead of milliseconds.
Another interesting modification could be combining the information obtained from the temperature compensation and the value from the synchronization algorithm to see if even better performance can be reached.

Of course, more accurate methods to better estimate the drift are available but are more computing intensive, because they do not use the sole time difference between two significant errors \cite{driftEstimation} \cite{driftEstimationPaper}. 
Nevertheless, even if the algorithm uses only few messages, it has been proven good enough for soft real time applications on low-resource sensors, that need synchronization errors in the order of the milliseconds and cannot afford to send large amount of synchronization messages or computing effort.


\balance

\section*{Acknowledgment}
The project presented in this paper has been funded with the contribution of the Autonomous Province of Trento, Italy, through the Regional Law 6/98.
Name of the granted Project: LT4.0. In addition, the authors gratefully acknowledge the support from the BLM Group.



%

\end{document}